\documentclass[twoside,reqno]{HERON}
\usepackage{epsfig,cite,colordvi}
\usepackage{graphicx}
\usepackage{url}
\usepackage{amssymb,amsmath,amscd,epsf}
\usepackage{times}
\usepackage{makeidx}
\newcommand{\pbard}{{\bar p} \, d \to e^+ e^- n}

\newcommand{\mee}{{\cal M}}
\newcommand{\panda}{\overline{\mathrm{P}} \mathrm{ANDA}}

\begin{document}

\title{Proton Timelike Form Factors Near Threshold via
Antiproton-Nucleus Electromagnetic Annihilation}

\runningheads{Proton Timelike Form Factors near threshold...}
{V.A.~Karmanov, H.~Fonvieille}

\begin{start}

\author{V.A.~Karmanov}{1}, \coauthor{H.~Fonvieille}{2}

\index{Karmanov, V.A.} \index{Fonvieille, H.}

\address{Lebedev Physics Institute, Leninsky prospekt 53, 119991 Moscow, Russia}{1}

\address{LPC, Universit\'e Blaise Pascal, IN2P3, 63177 Aubi\`ere Cedex, France}{2}

\begin{Abstract}
In the reaction of the antiproton-deuteron electromagnetic
annihilation $\bar{p} + d \to n e^+ e^-$ the value of the
invariant mass $M_{e^+ e^-}$ of the $e^+ e^-$ pair can be near or
below the $\bar{p} p$ mass even for enough high momentum of
incident antiproton. This allows to access the proton
electromagnetic form factors in the time-like region of $q^2$ near
the $\bar{p} p$ threshold. We estimate the cross section
$d\sigma(\bar{p} +d \to e^+ e^- n)/dM_{e^+ e^-}$ for an antiproton
beam momentum of 1.5 GeV/c. We find that for values of $M_{e^+
e^-}$ near the $\bar{p} p$ threshold this cross section is about 1
pb/MeV. The case of heavy nuclei $^{12}$C, $^{56}$Fe and
$^{197}$Au is also estimated. Elements of experimental feasibility
are studied in the context of the $\panda$ project
\cite{Panda:2009,Panda:2005}. We conclude that this process has a
chance to be measurable at $\panda$.
\end{Abstract}
\end{start}









\section{Introduction}
\label{intro}

The electromagnetic form factors of the proton and the neutron are
basic observables, which are the goal of  extensive measurements.
In the spacelike region, \ie for a virtual photon four-momentum
squared $q^2<0$, these form factors give information about the
spatial distribution of electric charge and magnetization inside
the nucleon. In the timelike region ($q^2>0$) they tell us about
the dynamics of the nucleon-antinucleon ($N \bar N$) interaction.

\begin{figure}[h!]
\begin{center}
\includegraphics[width=8cm]{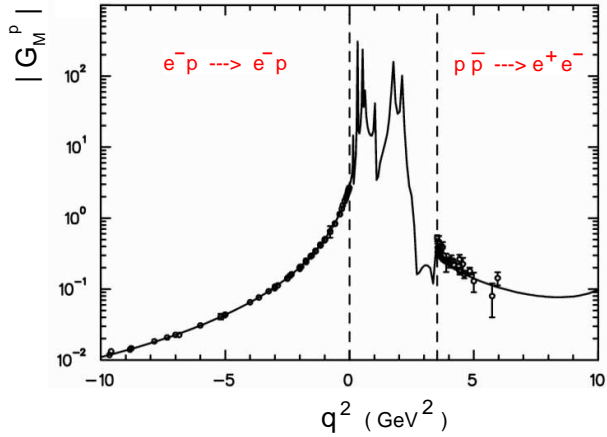}
\caption{Experimental data and predictions for the magnetic proton
form factor in the domain $-10\;GeV^2 \le q^2 \le 10\;GeV^2$. The
figure is taken from \cite{Meissner:1999hk}. \label{fig1}}
\end{center}
\end{figure}

The theoretical models, generally based on  dispersion relations
\cite{Baldini:1998qn,Hammer:1996kx,Adamuscin:2005aq} or
semi-phenomenological approaches
\cite{Iachello:2004aq,Bijker:2004yu}, predict a smooth behavior of
the form factor in the measured regions,  but a peaked behavior in
the timelike region below the $N \bar N$ threshold ($0 < q^2 <
4m^2$, $m$ is the nucleon mass), due to poles in the amplitude
(see \eg fig.~\ref{fig1}, taken from \cite{Meissner:1999hk}).
These poles are phenomenological inputs, built from meson
exchange,  and their properties are fitted to the data in the
measured regions. The corresponding irregularities in form factors
are related to the transition of $p\bar{p}$ to vector mesons which
can decay in $e^+e^-$ pair via a virtual photon.

The mesons with a  mass near the $p\bar{p}$ mass can have a
quasinuclear nature, \ie, they can be formed by bound states and
resonances in the $p\bar{p}$ system. Such vector mesons were
predicted in the papers~\cite{Dalkarov:1989xs,Dalkarov:1989dc}.
Note that such mesons can be formed not only in the $p\bar{p}$
system but in $N\bar{N}$ in general and they can  have not only
vector quantum numbers. A review on quasinuclear mesons in the
$N\bar{N}$ system is given in \cite{Shapiro:1978wi}.

The under-threshold region ($0 < q^2 < 4m^2$) is called unphysical
because it cannot be accessed experimentally by an on-shell
process. Some experiments have been performed in the vicinity of
the $N \bar N$ threshold, either in $p \bar p \to e^+ e^-$  at
LEAR \cite{Bardin:1994am} or in the inverse channel $e^+ e^- \to p
\bar p$ at Babar \cite{Aubert:2005cb}, but they cannot go below
this physical threshold. However, a nucleus provides nucleons with
various momenta, in modulus and direction, and also various
degrees of off-shellness. Therefore it offers the possibility to
produce an $N \bar N$ electromagnetic annihilation with an
invariant mass squared $q^2= s_{\bar p p}$ smaller than $4 m^2$.

This possibility, which may give access to the proton form factors
in the underthreshold region, for an off-shell nucleon, was
explored in our paper \cite{fk2009}. The present talk is based on
this paper.

The idea to use a nucleus for that purpose was explored in the
80's using deuterium ~\cite{Dalkarov:1980de}. The reaction is
then:
\begin{equation}\label{(1)}
\pbard
\end{equation}
(a crossed-channel of deuteron electrodisintegration). The aim of
the present paper is to revive this study in view of the future
antiproton facility FAIR at GSI.

Other channels can give access to the off-shell nucleon form
factors in the timelike region, including the underthreshold
region; such processes have been studied theoretically in
ref.~\cite{Schafer:1994vr} ($\gamma p \to p e^+ e^-$) and in
refs.~\cite{Adamuscin:2007iv,Dubnickova:1995ns}
 (${\bar p} p \to \pi^0 e^+ e^-$).

This paper is organized as follows: a theoretical study is
presented in sect. \ref{sec-globalth}, experimental aspects are
discussed in sect.~\ref{sec-globalexp} and a conclusion is given
in sect.~\ref{sec-concl}.

\section{Theoretical study}
\label{sec-globalth}

In elastic electron scattering from the nucleon  $e^-N\to e^-N$
the momentum transfer squared $q^2=(k-k')^2$ is always negative.
This allows to measure the nucleon form factors in the space-like
domain of $q^2$.

On the contrary, in the annihilation $N\bar{N}\to \gamma^*\to
e^+e^-$ the mass of virtual photon is equal to the total c.m.
$N\bar{N}$ energy. Its four-momentum squared is always greater
than $4m^2$. This allows to measure the nucleon form factors in
the time-like domain of $q^2$, above the $N\bar{N}$ threshold. In
this reaction, in order to study the form factor behavior in a
narrow domain near threshold, where non-trivial structures are
predicted \cite{Shapiro:1978wi}, one should have a beam of almost
stopped antiprotons. This non-easy technical problem was solved at
LEAR ~\cite{Bardin:1994am}. However, the under-threshold  domain
$0\le q^2 \le 4m^2$ remains kinematically unreachable in this type
of experiments.

\begin{figure}[h!]
\begin{center}
\includegraphics[width=6cm]{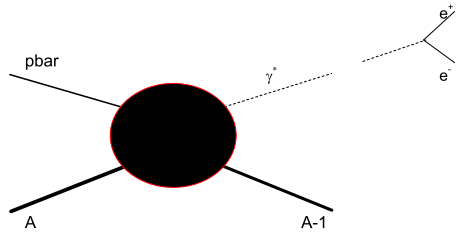}
\end{center}
\caption{The process $\bar{p}A\to (A-1)\gamma^*$  (followed by
$\gamma^*\to e^+e^-)$.} \label{pbarA}
\end{figure}

\begin{figure}[h!]
\begin{center}
\includegraphics[width=7cm]{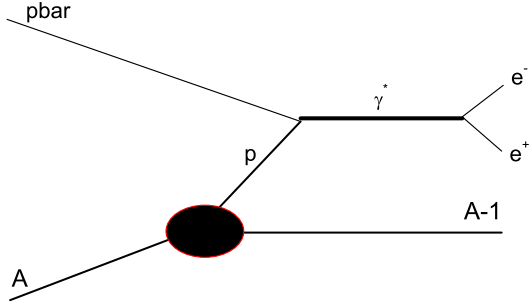}
\end{center}
\caption{Amplitude of the reaction $\bar{p}A\to (A-1)\gamma^*$ in
impulse approximation. } \label{eAMX}
\end{figure}

One can penetrate in this domain of $q^2$ in the $\bar{p}$
annihilation on nuclei
$$\bar{p}A\to (A-1) \ e^+e^-,$$
see fig. \ref{pbarA}. The symbol $(A-1)$ means not necessarily a
nucleus but any system with the baryon number $A-1$. Since extra
energy of the antiproton can be absorbed by the $(A-1)$ system,
the $e^+e^-$ pair may be emitted with very small invariant mass.
Therefore the two-body reaction $\bar{p}A\to (A-1)\gamma^*$ is
kinematically allowed for a very wide domain of invariant mass of
the $\gamma^*$, which starts with two times the electron mass,
namely:
$$4m_e^2\le q^2 \le (\sqrt{s_{\bar{p}A}}-M_{A-1})^2 \ . $$
One can achieve near-threshold, under-threshold and even
deep-under-threshold values of $q^2$ even for fast antiprotons.
This however does not mean that this reaction provides us direct
information about the nucleon form factors. For the latter, we
should be sure that the observed $e^+e^-$ pair (and nothing more)
was created in the annihilation $\bar{p}p\to e^+e^-$ on the proton
in the nucleus, \ie, that the reaction mechanism is given by the
diagram of fig. \ref{eAMX} or by a similar diagram where the
$\bar{p}$ can rescatter before annihilation.

At the same time, since the nucleons in the nucleus are
off-mass-shell, the form factors  entering the amplitude of fig.
\ref{eAMX}, are not precisely the same as found in the free
$\bar{p}p$ annihilation. In general, the three-leg vertex
$F=F(p^2_1,p^2_2,q^2)$ depends not only on the photon virtuality
$q^2$, but also on the nucleon ones $p^2_1,p^2_2$. In the case
considered, the incident antiproton is on-ener\-gy-shell:
 $p^2_{\bar{p}}=m^2$, however the form
factors depend on the proton virtuality $p^2_p$. How the form
factor $F(p^2_{\bar{p}}=m^2,p^2_p\neq m^2,q^2)$ \vs $q^2$ differs
from the free one $F(p^2_{\bar{p}}=m^2,p^2_p= m^2,q^2)$ -- this
depends on the dynamics determining its behavior \vs the nucleon
leg virtuality. The nucleon form factors with off-shell nucleons
were studied in the papers
\cite{Naus:1987kv,Tiemeijer:1990zp}.
Generally, we can expect that the form factor dependence \vs
$p^2_p$ is much smoother than the $q^2$ dependence. The $p^2_p$
dependence can be determined by the nucleon self-energy
corrections (\ie, by the structure of the nucleon), whereas the
$q^2$ dependence in the time-like domain is governed  by the
$\bar{p}p$ interaction. The nucleon dynamics has a much larger
energy scale than the nuclear one. The typical off-shell variation
found in the papers
\cite{Naus:1987kv,Tiemeijer:1990zp}
was from a few to 10 percent. We do not pretend to such an
accuracy here. Therefore we neglect this effect in our
calculation. We will come to this question later. In any case,
both domains: $q^2<4m^2,\; p^2_p=m^2$ and $q^2<4m^2,\; p^2_p<m^2$
are totally unexplored experimentally and are interesting and
intriguing.

We emphasize that though the form factor dependence on $p^2_p$ can
be weak, the nucleon off-mass-shell effect is very important for
the kinematical possibility to reach the near- and under-threshold
domain of $q^2$ with fast antiprotons. To produce the
near-threshold $e^+e^-$ pairs in annihilation of a fast $\bar{p}$
on an on-mass-shell proton, the antiproton should meet in the
nucleus a fast proton with parallel momentum. The probability of
that, which was estimated in the Glauber approach, is negligibly
small relative to the results presented below. However, if the
effective mass $p^2_p$ of the virtual proton is smaller than $m^2$
(that is just the case in a nucleus), then the near- and
under-threshold $e^+e^-$ pairs can be produced in collisions with
not so fast intra-nucleus nucleons. This effect considerably
increases the cross section. To have an idea of the order of
magnitude which one can expect for this cross section, we will
calculate it in the impulse approximation. Numerical applications
will be done for the lowest antiproton beam momentum foreseen in
future projects. Namely,  at the High Energy Storage Ring at
FAIR-GSI this value is 1.5 GeV/c.

\subsection{Cross section calculation}
\label{sec:theocross}

At first, we consider the case of the deuteron target. If we know
the amplitude of the reaction $\bar{p}d\to e^+e^-n$: $M_{\bar{p}d
\to e^+e^-n}$ (to be calculated below), then the corresponding
cross section is given by:
\begin{eqnarray}\label{dsig1}
&&d\sigma_{\bar{p}d\to e^+e^-n} =\frac{(2\pi)^4}{4I}|M_{\bar{p}d
\to e^+e^-n}|^2
\\
&&\times \delta^{(4)}(p_{\bar{p}}+p_d-p_{e^+}-p_{e^-}-p_n)
\frac{d^3 p_{e^+}}{(2\pi)^3 2\epsilon_{e^+}}\;\frac{d^3
p_{e^-}}{(2\pi)^3 2\epsilon_{e^-}}\; \frac{d^3 p_n}{(2\pi)^3
2\epsilon_n} \nonumber
\end{eqnarray}
where $I$ results from the flux factors. Here and below we imply
the sum over the final spin projections and average over the
initial ones.

Our estimations are carried out in the impulse approximation, when
the mechanism is given by the diagram of fig. \ref{eAMX}. Then the
total amplitude squared $|M_{\bar{p}d \to e^+e^-n}|^2$ is
proportional to the annihilation amplitude squared $|M_{\bar{p}p
\to e^+e^-}|^2$ and to the square of the deuteron wave function
$|\psi|^2$:
\begin{equation}\label{Mbar}
|M_{\bar{p}d \to e^+e^-n}|^2 = 4m\,|M_{\bar{p}p \to
e^+e^-}|^2\;|\psi|^2, \
\end{equation}
and $|\psi|^2$ is normalized to 1.

We are interested in the distribution  in the invariant mass
${\cal M}$ of the final $e^+e^-$ system. To find it, for fixed
value of ${\cal M}$, we can integrate, in some limits, over the
angles of the recoil neutron (determining the neutron recoil
momentum) and over the angles of the emitted $e^+e^-$ in their
center of mass. This can be done using standard phase volume
techniques. The derivation is presented in detail in
\cite{fk2009}. The final result reads:
\begin{equation}\label{sig2}
\frac{d\sigma_{\bar{p}d \to e^+e^-n}}{d\mee}=\sigma_{\bar{p}p \to
e^+e^-}(\mee)\;\eta(\mee),
\end{equation}
where $\eta(\mee)$ is  the distribution (given by eq. (\ref{sig3})
below) of the $e^+e^-$ invariant mass $\mee$ and $\sigma_{\bar{p}p
\to e^+e^-}(\mee)$ is the cross section of the $\bar{p}p \to
e^+e^-$ annihilation at the total energy $\mee$.

The calculation of $\sigma_{\bar{p}p \to e^+e^-}$ is standard. To
estimate the nuclear effect, we omit the nucleon electromagnetic
form factors. Then the $\bar{p}p \to e^+e^-$ cross section obtains
the form:
\begin{equation}\label{sig4}
\sigma_{\bar{p}p \to e^+e^-}
=\frac{2\alpha^2\pi(2m^2+\mee^2)}{3\mee^2m p_{\bar{p},lab}} \ .
\end{equation}

Though $\sigma_{\bar{p}p \to e^+e^-}$ depends on $\mee$, the main
(nuclear) effect is determined by the factor $\eta(\mee)$
\cite{fk2009}:
\begin{eqnarray}\label{sig3}
\eta(\mee)&=&\frac{m p^*_{\gamma^*
n}\mee}{(2\pi)^2\sqrt{s_{\bar{p}d}}}\int_{-1}^1|\psi(k)|^2dz.
\end{eqnarray}
Here $z=\cos\theta$, where $\theta$ is the angle, in the c.m.
frame of the reaction, between the initial deuteron momentum
$\vec{p}^*_d$ and the final neutron momentum $\vec{p}^*_n$. The
argument of the wave function $k$ depends on $z$. This explicit
dependence is given in \cite{fk2009}.

Since, as mentioned, the wave function squared $|\psi(k)|^2$ is
normalized to 1, the distribution $\eta(\mee)$ is also
automatically normalized to 1:
\begin{equation}\label{norm}
\int_0^{\infty}\eta(\mee)d \mee= 1.
\end{equation}

The total cross section is obtained by integrating (\ref{sig2}) in
the finite limits $\mee_{min}\le \mee\le \mee_{max}$, where
$\mee_{min}=2m_{e}$, $\mee_{max}=\sqrt{s_{\bar{p}d}}-m$.
Neglecting the electron mass, we can put $\mee_{min}=0$. For
$p_{\bar{p},lab}=1.5$ GeV/c the value $\mee_{max}$ is high enough
and provides the normalization condition (\ref{norm}) with very
high accuracy.

To emphasize more distinctly the effect of the nuclear target, we
can represent the cross section (\ref{sig4}) of the annihilation $
\bar{p}p \to e^+e^-$ on a free proton similarly to eq.
(\ref{sig2}):
\begin{equation}\label{sig5}
\frac{d\sigma_{\bar{p}p \to e^+e^-}}{d\mee}=\sigma_{\bar{p}p \to
e^+e^-}\delta(\mee-\sqrt{s_{p\bar{p}}})
\end{equation}
where $\sigma_{\bar{p}p \to e^+e^-}$ is defined in (\ref{sig4}),
$s_{p\bar{p}}=(p_p+p_{\bar{p}})^2$. The fact that in the
annihilation on a free proton the mass of the final $e^+e^-$ pair
is fixed is reflected in (\ref{sig5}) in the presence of the
delta-function. Comparing this formula with (\ref{sig2}), we see
that the effect of the nuclear target results in a dilation of the
infinitely sharp distribution $\delta(\mee-\sqrt{s_{p\bar{p}}})$
in a distribution of finite width $\eta(\mee)$. The dilation of a
distribution does not change its normalization: $\eta(\mee)$
remains normalized to 1.

\subsection{Analysis and numerical calculations}
\label{sec:theo-ana-num}

At first glance, the small near-threshold $\bar{p}p$ c.m. energy
$\mee\approx 2m$ in the collision of a fast $\bar{p}$
($p_{\bar{p}}=1500$ MeV/c) is achieved, when the antiproton meets
in the deuteron a fast proton having the same momentum as the
$\bar{p}$, in modulus and direction. The protons with such a high
momentum are very seldom in deuteron. For this mechanism, the
cross section would be very small. However, the near-threshold
value of $\mee$ is obtained in other kinematics. As  we mentioned,
the proton momenta $k$ in the deuteron wave function $\psi(k)$ in
eq. (\ref{sig3}) start with $k\approx k_{min}\approx 360$ MeV/c
only (that corresponds to  $z\approx 1$). The main reason which
allows to obtain in this collision the value $\mee\approx 2m$ is
the off-shellness of the proton: $m^*\le 0.85 m$ instead of
$m^*=m$. This 15\%  decrease relative to the free proton mass is
enough to obtain the invariant $p\bar{p}$ mass $\mee\approx 2m$,
when one has the two parallel momenta: 1500 MeV/c for $\bar{p}$
and 360 MeV/c for $p$.

\begin{figure}[h!]
\begin{center}
\includegraphics[width=8cm]{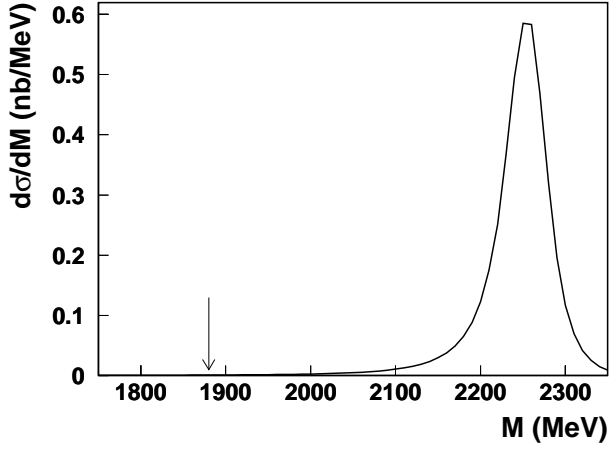}
\end{center}
\caption{The cross section $\frac{d\sigma_{\bar{p}d \to
e^+e^-n}}{d\mee}$ of the reaction $\bar{p}d\to e^+e^- n$ \vs
$\mee$, in the interval: $1750\;MeV \leq \mee \leq 2350 \;MeV$,
calculated for a pointlike proton. The arrow indicates the $p \bar
p$ threshold.} \label{newplot2}
\end{figure}

The cross section $d\sigma_{\bar{p}d \to e^+e^-n}/d\mee$, eq.
(\ref{sig2}), has been calculated  for an antiproton of momentum
$p_{\bar{p}}=1500$ MeV/c on a deuteron nucleus at rest, with the
deuteron wave function \cite{Carbonell:1995yi}, incorporating two
components corresponding to S- and D-waves. The result is shown in
fig. \ref{newplot2}. The maximum of the cross section is at
$\mee=2257$ MeV, that corresponds to the $\bar{p}$ interacting
with a proton at rest (and on-shell).
The cross section integrated over $\mee$ is equal to 43 nb. We
remind that these calculations do not take into account the proton
form factor. Its influence will be estimated below. The numerical
integral over $\mee$ of the function $\eta(\mee)$, eq.
(\ref{sig3}), is $\approx 1$, in accordance with the normalization
condition (\ref{norm}).

The $p \bar p$ threshold value $\mee= 1880$ MeV is on the tail of
the distribution, far from the maximum. Relative to the maximum,
the cross section at threshold decreases approximately by a factor
600. The numerical value at the threshold is:
\begin{eqnarray}
\left.\frac{d\sigma}{d\mee}\right|_{\mee=2m}=1\; \frac{pb}{MeV} \
. \label{hf1pb}
\end{eqnarray}

\begin{figure}[h!]
\begin{center}
\includegraphics[width=8cm]{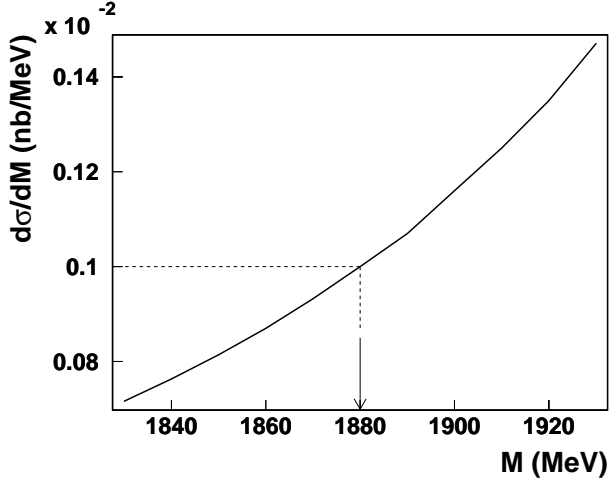}
\end{center}
\caption{The same as in fig.~\ref{newplot2}, but near the
$p\bar{p}$ threshold, in the interval $1830\;MeV \leq \mee \leq
1930 \;MeV$.}
 \label{newplot3}
\end{figure}

In fig.~\ref{newplot3} this cross section is shown in the
near-threshold interval $1830\;MeV \leq \mee \leq 1930 \;MeV$. The
integral over $\mee$ in a bin of width 100 MeV centered on the
threshold is:
$$
\int_{1830\;MeV}^{1930\;MeV}\frac{d\sigma_{\bar{p}d\to e^+e^-
n}}{d\mee}\;{d\mee}\approx 100\;pb \ .
$$

These estimations take into account the suppression resulting from
the momentum distribution in deuteron. However, they do not
incorporate the form factors of the nucleon. To incorporate them
in a simplified way, one can consider an effective form factor
$\vert F \vert$ which depends on $\mee$, and include it in the
integral:
\begin{equation}
\sigma_{\bar{p}d\to e^+e^-n} = \int  \sigma_{\bar{p}p \to
e^+e^-}(\mee)\;\eta(\mee) \; \vert F(\mee) \vert^2 d \mee
\label{cswithff}
\end{equation}
where $\sigma_{\bar{p}p \to e^+e^-}(\mee)$ is the cross section
for pointlike nucleons given in eq.(\ref{sig4}). To have an
estimate of this  integral, we have taken the effective proton
form factor  measured in ref.~\cite{Aubert:2005cb}. By doing this,
we neglect all off-shell effects.  We interpolate $\vert F(\mee)
\vert$ linearly between the measured values, and we limit the
integral to the region $\mee \ge 2m$. In this way we obtain
$\sigma_{\bar{p}d\to e^+e^-n} \simeq$ 1 nb, which is comparable to
the total cross section  $\sigma_{\bar{p}p \to e^+e^-}$ on a free
proton at $\mee=2257$ MeV. We point out that at threshold, our
differential cross section $ d\sigma_{\bar{p}d \to n e+e-}$ of 1
pb/MeV (eq.(\ref{hf1pb})) is not suppressed by any factor, since
there the form factor $\vert F \vert$ seems to be close to 1
experimentally \cite{Baldini:2007qg,Baldini:2008nk}.  Below this
threshold one may expect a form factor effect {\it larger} than
one.

\subsection{Annihilation on heavier nuclei}
\label{sec:heavier-nuclei}

For $A>2$, we should  take into account the possibility of
excitation and breakup of the final nucleus $A-1$ in the process
$\bar{p}A\to (A-1)\gamma^*$. The result contains the sum over the
final energies of the residual nucleus and the integral over a
continuous spectrum. That is, the function $|\psi(k)|^2$ in eq.
(\ref{sig3}) is replaced by the integral
$\int_{E_{min}}^{E_{max}}S(E,k)dE$, where $S(E,k)$ is the nucleus
spectral function giving the probability to find in the final
state the nucleon with the relative momentum $k$ and the residual
nucleus with energy $E$. For high incident energy we can replace
the upper limit by infinity. Then we obtain:
$$
\int_{E_{min}}^{\infty}S(E,k)dE=n(k),
$$
where $n(k)$ is the momentum distribution in the nucleus.

To estimate the cross section  on heavy nuclei, we will still use
eqs. (\ref{sig2}), (\ref{sig3}) but with the two following
changes. ({\it i}) We replace the deuteron momentum distribution
by the nuclear one. ({\it ii}) We multiply (\ref{sig3}) by the
number of protons $Z$.

The numerical calculations were carried out for the $^{12}$C,
$^{56}$Fe and $^{197}$Au nuclei with the nuclear momentum
distributions found in the papers by A.N.~Anto\-nov \etal:
\cite{Antonov:2004xm,Antonov:2006md}.
Near threshold, \ie at $\mee=1880$ MeV, for all three nuclei we
obtain very close results given by (compare with eq. (\ref{hf1pb})
for deuteron):
\begin{equation}\label{nucl}
\frac{d\sigma_{\bar{p}A\to e^+e^- X}}{d\mee}\; \approx
6.5\,Z\,\frac{pb}{MeV}
\end{equation}
Multiplying by the charge $Z$ ($Z(^{12}$C$  )=6 $,
 $Z(^{56}$Fe$ )=26$,
 $Z(^{197}$Au$)=79$)
and integrating (\ref{nucl}) over a 1 MeV interval near
$\mee=1880$ MeV, we get:

$$
\sigma(^{12}\mbox{C})=39 \;pb,\;\;
\sigma(^{56}\mbox{Fe})=0.17\;nb, \;\;
\sigma(^{197}\mbox{Au})=0.5\;nb.
$$

These results were obtained without taking into account the
absorption of $\bar{p}$ in nucleus before electromagnetic
annihilation. This absorption was estimated in Glauber approach.
It reduces these cross sections by only a factor 2.


\subsection{Beyond the impulse approximation}
\label{sec:beyond} There exist other possible mechanisms for the
process $\pbard$. One of them is the initial state interaction,
which includes rescattering (not only elastic) of the initial
$\bar{p}$ in the target nucleus. In the rescattering, the incident
$\bar{p}$ looses energy and therefore the proton momentum needed
to form the invariant mass $\mee\approx 2m$ becomes smaller. The
probability to find such a proton in deuteron is higher. Therefore
initial state interaction increases the cross section.

Other processes are discussed in \cite{fk2009}. We emphasize that
in any case, whatever the intermediate steps are in process
(\ref{(1)}), the $e^+e^-$ pair of the final state must come
necessarily from the baryon-antibaryon electromagnetic
annihilation, $\bar{p}p$ or $\bar{n}n$, because there is only one
neutron left at the end. It cannot come from another process, even
if there are complicated intermediate steps, like rescattering,
etc. Therefore this $e^+e^-$ pair is a direct and very little
distorted probe of the baryon-antibaryon electromagnetic
annihilation vertex.

\section{Experimental aspects}
\label{sec-globalexp}

Experimental aspects have been investigated in detail in
\cite{fk2009} in the case of a deuteron target, \ie for the
three-body process $\pbard$, in the conditions of the $\panda$
project at FAIR-GSI: an antiproton beam momentum of 1.5 GeV/c and
the detection of the lepton pair. The count rate in the
near-threshold region of $\mee$ is small but not negligible. The
main difficulty is to identify the reaction among the hadronic
background which is about six orders of magnitude higher. First
elements of strategy were presented for this background rejection,
based on particle identification, detector hermeticity, and
missing mass resolution.

One should also also note that the luminosity in ($\bar{p}$A)
decreases with the atomic charge $Z$ of the target nucleus
\cite{Panda:2009}, in a way that roughly compensates the increase
of cross section with $Z$ reported in sect.
\ref{sec:heavier-nuclei}.

Although the subject would require a much more detailed study, it
was concluded that this process has a chance to be measurable in
$\panda$, given the very good design performances of the detector.

The antiproton momentum 1.5 GeV/c just corresponds to the
threshold value of creation of the $\Lambda\bar{\Lambda}$ pair on
a free proton. Therefore the virtual creation of the
$\Lambda\bar{\Lambda}$ pair in reaction (\ref{(1)}) is not
suppressed by the nucleon momentum distribution in deuteron and
contributes just in the domain of the peak of fig.~\ref{newplot2},
that allows one to study the $\Lambda\bar{\Lambda}$ threshold
region with good statistics. In the $\Lambda\bar{\Lambda}$ system,
the quasi-nuclear states were predicted in \cite{Carbonell:1993dt}
and, similarly to the $N\bar{N}$ quasi-nuclear states, they should
manifest themselves as irregularities in the cross section. The
contribution of the channel $\bar{p}p\to \bar{\Lambda}\Lambda\to
e^+e^-$ in the total cross section $\bar{p}p\to e^+e^-$ (the
latter equals 1 nb, see sect.~\ref{sec:theo-ana-num} above) was
estimated in~\cite{Dalkarov:private} as 0.1 nb, \ie 10\% of the
total cross section. Therefore we expect that the structures
caused by the channel $\bar{p}p\to \bar{\Lambda}\Lambda\to e^+e^-$
can be observed in process (1) in the region of mass $\mee$ near
the $\Lambda\bar{\Lambda}$ threshold.

\section{Conclusion}
\label{sec-concl}

We have studied the reaction  $\bar{p}A\to (A-1)\gamma^*$
(followed by $\gamma^*\to e^+e^-)$. This process gives access to
the ${\bar p} p$ annihilation $\bar p p \to \gamma^*$ at invariant
masses $\sqrt s_{{\bar p} p}$ which are below the physical
threshold of  $ 2 m$, due to the proton off-shellness in the
nucleus. In this way a possibility exists to access the proton
timelike form factors in the near-threshold and the totally
unexplored under-threshold region, where  $N \bar N$ bound states
are predicted.

The differential cross section $d\sigma / d \mee $ has been
calculated as a function of the dilepton invariant mass $\mee$,
for an incident antiproton of 1.5 GeV/c momentum on a deuteron
target (and heavier nuclei). We find that the $p \bar p$ threshold
($\mee=2m$) is reached for a minimal proton momentum
$k_{\min}$=360 MeV/c in the nucleus, and at this point the cross
section is about 1 pb/MeV.

The estimations carried out above show that this process has a
chance to be measurable in $\panda$, provided the very good design
performances of the detector.

\end{document}